\documentclass[amsmath,nobibnotes,aps,superscriptadress,amssymb,pra,aps,showpacs,superscriptaddress,twocolumn]{revtex4-1}

\usepackage{graphicx}
\graphicspath{{./}{./figs/}}
\usepackage{color}
\usepackage{makecell}
\usepackage[colorlinks, linkcolor=blue, urlcolor=blue, citecolor=blue]{}
\usepackage{lettrine}
\usepackage[normalem]{ulem}
\usepackage{mathtools} 
\usepackage{upgreek}
\usepackage{ragged2e}
\usepackage{pgfplotstable}
\usepackage{catchfilebetweentags}
\usepackage{chngcntr}

\usepackage{amsmath}
\makeatother
\newcommand{\beq}{\begin{equation}}
\newcommand{\eeq}{\end{equation}}
\newcommand{\beqa}{\begin{eqnarray}}
\newcommand{\eeqa}{\end{eqnarray}}

\begin{document}
\title{A Neural Network Assisted  $^{171}$Yb$^{+}$ Quantum Magnetometer}
\author{Yan Chen}
\affiliation{CAS Key Laboratory of Quantum Information, University of Science and Technology of China, Hefei 230026, China}
\affiliation{CAS Center For Excellence in Quantum Information and Quantum Physics, University of Science and Technology of China, Hefei 230026, China}

\author{Yue Ban}
\email{ybanxc@gmail.com}
\affiliation{Department of Physical Chemistry, University of the Basque Country UPV/EHU, Apartado 644, 48080 Bilbao, Spain}
\affiliation{EHU Quantum Center, University of the Basque Country UPV/EHU, Barrio Sarriena, s/n, 48940 Leioa, Biscay, Spain}
\affiliation{TECNALIA, Basque Research and Technology Alliance (BRTA), 48160 Derio, Spain}

\author{Ran He}
\email{heran@mail.ustc.edu.cn }
\affiliation{CAS Key Laboratory of Quantum Information, University of Science and Technology of China, Hefei 230026, China}
\affiliation{CAS Center For Excellence in Quantum Information and Quantum Physics, University of Science and Technology of China, Hefei 230026, China}

\author{Jin-Ming Cui}
\email{jmcui@ustc.edu.cn}
\affiliation{CAS Key Laboratory of Quantum Information, University of Science and Technology of China, Hefei 230026, China}
\affiliation{CAS Center For Excellence in Quantum Information and Quantum Physics, University of Science and Technology of China, Hefei 230026, China}
\affiliation{Hefei National Laboratory, University of Science and Technology of China, Hefei 230088, China}

\author{\\Yun-Feng Huang}
\email{hyf@ustc.edu.cn}
\affiliation{CAS Key Laboratory of Quantum Information, University of Science and Technology of China, Hefei 230026, China}
\affiliation{CAS Center For Excellence in Quantum Information and Quantum Physics, University of Science and Technology of China, Hefei 230026, China}
\affiliation{Hefei National Laboratory, University of Science and Technology of China, Hefei 230088, China}

\author{Chuan-Feng Li}
\email{cfli@ustc.edu.cn}
\affiliation{CAS Key Laboratory of Quantum Information, University of Science and Technology of China, Hefei 230026, China}
\affiliation{CAS Center For Excellence in Quantum Information and Quantum Physics, University of Science and Technology of China, Hefei 230026, China}
\affiliation{Hefei National Laboratory, University of Science and Technology of China, Hefei 230088, China}

\author{Guang-Can Guo}
\affiliation{CAS Key Laboratory of Quantum Information, University of Science and Technology of China, Hefei 230026, China}
\affiliation{CAS Center For Excellence in Quantum Information and Quantum Physics, University of Science and Technology of China, Hefei 230026, China}
\affiliation{Hefei National Laboratory, University of Science and Technology of China, Hefei 230088, China}

\author{Jorge Casanova}
\affiliation{Department of Physical Chemistry, University of the Basque Country UPV/EHU, Apartado 644, 48080 Bilbao, Spain}
\affiliation{EHU Quantum Center, University of the Basque Country UPV/EHU, Barrio Sarriena, s/n, 48940 Leioa, Biscay, Spain}
\affiliation{IKERBASQUE, Basque  Foundation  for  Science, Plaza Euskadi 5, 48009 Bilbao, Spain}

\begin{abstract}
A versatile magnetometer must deliver a readable response when exposed to target fields in a wide range of parameters. In this work, we experimentally demonstrate that the combination of $^{171}$Yb$^{+}$ atomic sensors with adequately trained neural networks enables to investigate target fields in distinct challenging scenarios. In particular, we characterize radio frequency (RF) fields in the presence of large shot noise, including the limit case of continuous data acquisition via single-shot measurements. Furthermore, by incorporating neural networks we significantly extend the working regime of atomic magnetometers into scenarios in which the RF driving induces responses beyond their standard harmonic behavior. Our results indicate the benefits to integrate neural networks at the data processing stage of general quantum sensing tasks to decipher the information contained in the sensor responses.
 \end{abstract}

\maketitle

\section*{Introduction}
\lettrine{Q}{uantum} sensing~\cite{Q-sensing} and metrology~\cite{Q-metrology} are important branches of modern quantum technologies with applications in different areas such as imaging~\cite{imaging1,imaging2} and spectroscopy~\cite{spectroscopy1,spectroscopy2,spectroscopy3}. In this scenario, atomic-sized sensors encoded in $^{171}$Yb$^{+}$~\cite{Yb1,Yb2,Yb3,Yb4,Yb5} and $^{40}$Ca$^{+}$~\cite{Ca} ions provide spatial resolution and sensitivity for the detection of external/target fields. In addition, $^{171}$Yb$^{+}$ ions exhibit negligible emission rates~\cite{Yb4} and extended coherence times due to the stabilization provided by dynamical decoupling (DD) techniques~\cite{DD1,DD2, Munuera-Javaloy21}. In particular, DD methods that exploit the multi-level structure in the $^2 S_{\frac{1}{2}}$ manifold of  the $^{171}$Yb$^{+}$ ion have lead to detection of radio frequency (RF) fields with sensitivity close to the standard quantum limit~\cite{Yb2}, while the resulting  dressed state qubit was proposed as a robust register for quantum information processing~\cite{Yb1,Yb3}. However, this approach is restricted to a narrow working regime leading to harmonic sensor responses where RF target parameters are encoded. A departure from such range leads to complex sensor responses where standard inference of the external fields gets challenging. In another vein, machine learning (ML) tools are  incorporated to address distinct problems in quantum technologies. In particular, neural networks (NNs) are valuable in distinct quantum sensing scenarios leading to adaptive protocols for phase estimation \cite{phase-estimation-ML1,phase-estimation-ML2,phase-estimation-ML3}, parameter estimation \cite{parameter-estimation-ML1,parameter-estimation-ML2,parameter-estimation-ML3,parameter-estimation-ML4,parameter-estimation-ML5}, and quantum sensors calibration \cite{calibration-ML1,calibration-ML2, sensor-NN}.

In this article, we experimentally demonstrate the ability of NNs to decode complex sensor responses, thus significantly extending the operational regime of quantum detectors. In particular, we infer RF target parameters from the measured response of an $^{171}$Yb$^+$ atomic sensor in distinct challenging scenarios. These comprise the regime of  non-harmonic sensor responses with a large shot noise, and the continuous interrogation of the sensor (via single-shot measurements) under always-on RF fields. With a careful modelling of the sensor-target dynamics, we train NNs to relate intricate responses of the sensor with RF parameters leading to estimations of the latter with high accuracy.

\section*{Results}
\subsection*{The quantum sensor} 
\begin{figure*}[t]
\begin{center}
\scalebox{0.29}[0.29]{\includegraphics{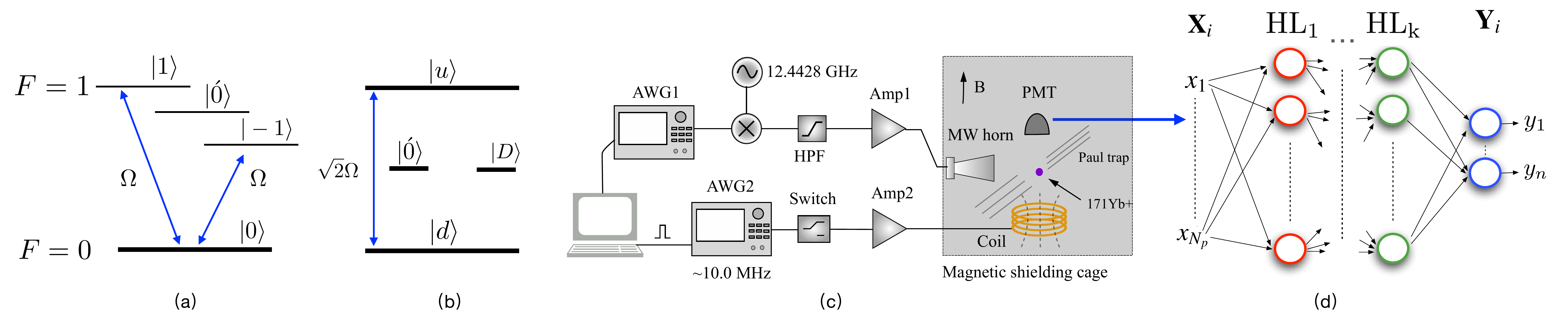}}
\caption{\label{Schematic-configuration} Schematic levels of the ${^{171}\rm Yb^{+}}$ atomic sensor and the experimental setup. (a) Relevant levels of the ${^{171}\rm Yb^{+}}$ atomic sensor. Two resonant MW fields drive the sensor with a Rabi frequency $\Omega$  leading to the configuration in (b) which is defined in the dressed state basis $\{|u\rangle$, $|d\rangle$, $|\acute{0}\rangle$, $|D\rangle \}$~\cite{Yb2}. (c) Schematic configuration of the experimental setup. The ion is trapped in a needle trap which consists of a pair of RF electrodes and four DC electrodes. The MW fields for state manipulation are generated by mixing a 12.4428 GHz signal with the signal from an arbitrary waveform generator (AWG1). The 12.6 GHz signal after a high pass filter (HPF) is amplified by an amplifier (Amp1) and sent to the ion using a MW horn outside the vacuum chamber. The target field around 10 MHz is generated by another arbitrary waveform generator (AWG2) controlled by the computer and broadcast to the ion through a coil after amplification. We place an extra RF switch after AWG2, which can be used to remove the target field from the trap while keeping the signal from AWG2 continuous. A photomultiplier tube (PMT) is used to detect the state-dependent fluorescence. (d) Scheme of the NN. The response from PMT, i.e. $\textbf{X}$, is processed by a number of hidden layers ($\rm{HL}_1$ ... $\rm{HL}_k$) leading to the outputs $\textbf{Y}$.  }
\end{center}
\end{figure*}

The sensor consists of four hyperfine levels ($|0\rangle, |\acute 0\rangle, |1\rangle$, and $|-1\rangle$) in the $^2 S_{\frac{1}{2}}$ manifold of the $^{171}$Yb$^{+}$ ion, and in a static magnetic field $B_z$. To enhance the coherence of the sensor, we use two microwave (MW) fields with Rabi frequency $\Omega$ leading to the dressed basis set $\{|u\rangle, |d\rangle, |D\rangle, |\acute{0}\rangle\}$ with $|u\rangle = (|B\rangle + |0\rangle) / \sqrt{2}$, $|d\rangle =  (|B\rangle-|0\rangle)/ \sqrt{2}$,  $|D\rangle =(|-1 \rangle -|1\rangle)/\sqrt{2}$ and $|B\rangle =(|-1 \rangle +|1\rangle)/\sqrt{2}$ which is insensitive  to magnetic field fluctuations \cite{Yb1,Yb2}, see Fig.~\ref{Schematic-configuration} (a) and (b).

Once a target field $\Omega_{\rm{tg}}\cos(\omega_{\rm{tg}}t +\phi_{\rm{tg}})$ is applied, one measures the dark state ($|D\rangle$) survival probability $P_D(t)$ considered here as the sensor response. The standard working regime of the sensor (leading to harmonic responses) implies $\Omega_{\rm{tg}} \ll \Omega  \ll \omega_{\rm{tg}}$ with $ \omega_{\rm{tg}}$ resonant with, e.g., the $|\acute0\rangle \leftrightarrow |1\rangle$ transition~\cite{Yb2}. This  simplifies the total Hamiltonian (see Supplementary Note 1) into 
\begin{equation}\label{Hharmonic}
H = -\frac{\Omega_{\rm{tg}}}{2\sqrt{2}} (|D\rangle \langle \acute{0}| +|\acute{0}\rangle \langle D| ).
\end{equation}
From Eq.~(\ref{Hharmonic}) one finds the harmonic sensor response $P_D(t) = \cos^2(\uppi t/t_{\rm{R}})$, with $t_{\rm{R}} =  2\uppi \sqrt{2}/ \Omega_{\rm{tg}}$ establishing a simple relation between $t_{\rm{R}}$ and the target field parameter $\Omega_{\rm{tg}}$. In contrast, a departure from the standard working regime leads to the loss of this type of simple  dependencies between sensor responses and target parameters, thus posing serious challenges for reliable field characterization. We demonstrate that our setup combining a quantum sensor and NNs can extract RF fields parameters from complex sensor responses.  

\subsection*{Experimental setup} 
As schematically shown in Fig.~\ref{Schematic-configuration}~(c), our protocol is executed on a single $^{171}\rm Yb^{+}$ ion confined in a Paul trap which is shielded by permalloy to reduce the surrounding magnetic noise~\cite{setup1,setup2}. A magnetic field $B_z$ is applied to the ion leading to a Zeeman shift $\approx 10.0$ MHz between $|\acute{0}\rangle$ and $|1\rangle$. Two $\sim$12.6 GHz MW fields $\Omega_j \cos{(\omega_j t +\phi_j)}$ $(j =1, 2)$ respectively resonant with transitions $|0\rangle \leftrightarrow |1\rangle$ and $|0\rangle \leftrightarrow |-1\rangle$ are used for state manipulation. To generate the dark state $|D\rangle$, we design pulses with the amplitudes $\Omega_{1}$ and $\Omega_{2}$ evolving in the form of a hyperbolic tangent (details available in the Supplementary Information). During the sensing window, the amplitudes of dressing fields are kept constant at $\Omega_1 = \Omega_2 = (2\uppi)\times5.5$ kHz. Afterwards the state remaining in $|D\rangle$ is transferred to $|0\rangle$ for detection.

A copper coil placed under the trap generates the target field once an RF current is sent to the coil.  As the original RF signal is produced by an arbitrary waveform generator (AWG2), the amplitude, frequency, and the initial phase of the target field can be set by adjusting the parameters of AWG2. An RF switch is placed after the AWG2 output in order to remove interaction between  the field and the ion if necessary while keeping the signal source continuously on. After amplification, the RF target field generated by the coil has a maximum amplitude of ${\Omega_{\rm{tg}}^{\rm{max}}} = (2\uppi)\times9.0$ kHz. A photomultiplier tube (PMT) is used to detect the state-dependent fluorescence.
As schematically shown in Fig. \ref{Schematic-configuration} (d), the response from PMT, i.e. $\textbf{X}$, is processed through the well-trained NN leading to the outputs $\textbf{Y}$ which approach the targets $\textbf{A}$. We use data generated from numerical simulations to create the NN. Then, by inputting experimentally collected responses into the NN, we estimate the RF parameters.

Now, we demonstrate the good performance of an $^{171}$Yb$^{+}$-magnetometer assisted by a NN in two scenarios in which  input data comprises  $(i)$ average values obtained from a reduced number of measurements, and  $(ii)$ binary sequences (including only 0, or 1) continuously acquired from single-shot measurements. Both cases are demonstrated in regimes where the sensor delivers complex responses that depart from the standard harmonic regime. Hence, we prove that NNs significantly extends the versatility of quantum sensors.

 \begin{figure*}[t]
\centering
\scalebox{0.22}[0.22]{\includegraphics{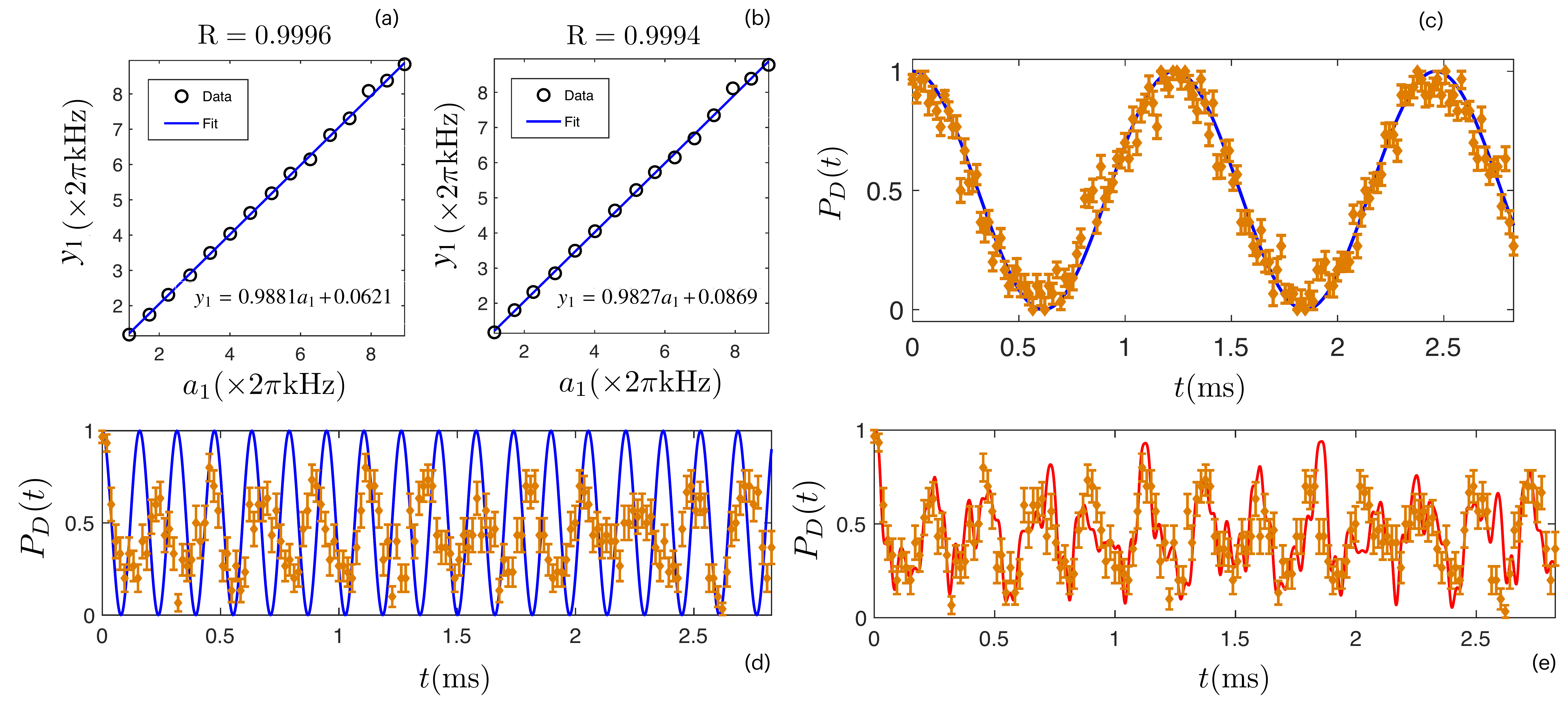}}
\caption{\label{average-times} Estimation results from Scenario i, a reduced number of measurements. (a-b) Regression of the NN outputs ($y_1$) with respect to the targets $a_1 \equiv \Omega_{\rm tg}$ for a number of shots $N_m=100$ (a) and $N_m=30$ (b). The fit (solid-blue) overlaps with the line $y_1 = a_1$, while R is the correlation coefficient~\cite{Correlation-coefficient} of $y_1$ and $a_1$. This is $\rm{R} = 0.9996$ (a), and  $0.9994$ (b). (c-e)  Sensor response (diamonds) for the targets $\Omega_{\rm tg} = (2\uppi) \times 1.1487$ kHz (c) and $\Omega_{\rm tg} = (2\uppi) \times 8.9493$ kHz (d, e) experimentally obtained for $N_m=30$ shots. For comparison, in (c,d) the harmonic response $P_D(t) = \cos^2(\uppi t/t_{\rm{R}})$  (solid-blue) is included. The plot in (c) shows the case with $\Omega_{\rm tg} = (2\uppi) \times 1.1487$ kHz lying in the harmonic regime, while in (d) the response for $\Omega_{\rm tg} = (2\uppi) \times 8.9493$ kHz significantly deviates from the harmonic behavior. In (e) we observe that the curve obtained from numerical simulations (solid-red) fits  experimental data for $\Omega_{\rm tg} = (2\uppi) \times 8.9493$ kHz. Error bars in (c-e) represent the standard error of the mean.}
\end{figure*}

\subsection*{Scenario i: Parameter estimation with a reduced number of  measurements} 
We aim to estimate the Rabi frequency of a target field. In this case, the input data string $\textbf{X}=\{P_1, P_2, ..., P_{N_p}\}$ consist on the average values $P_i$ ($i\in[1, N_p]$) collected at $N_p$ time instants $t = t_i$ distributed in a time interval $[0, t_f]$ for a specific $\Omega_{\rm{tg}}$.  We use the Hamiltonian in the Supplementary Information to numerically compute $P_D(t_i)$, as each $P_i$ does not follow the expression $P_D(t_i) = \cos^2(\uppi t_i/t_{\rm{R}})$ in cases that depart from the standard working regime. In addition, the binary outcome is drawn from a Bernoulli distribution $z_n^i \sim B(1, P_D(t_i)) \in\{0, 1\}$ such that each simulated average value is $P_i  = \sum_{n=1}^{N_m} z_n^i / N_m$ for a number of shots $N_m$. We choose $N_p = 151$, $N_m=100$ and $t_f = 2.828$ ms (note that this value of $t_f$ corresponds to one period of the sensor response for $\Omega_{\rm{tg}} = (2\uppi) \times 0.5$ kHz, i.e. in the harmonic case). The examples (i.e. the data strings \textbf{X}, $\textbf{Y}$, and $\textbf{A}$) are computed by selecting $96$ values for $\Omega_{\rm{tg}}/(2\uppi)$ in the range $[0.5, 10]$ kHz. In addition, as each $P_i$ fluctuates owing to the reduced number of measurements, we perform $100$ repetitions for each simulated experimental acquisition (i.e. for each $\Omega_{\rm{tg}}$). Therefore, our dataset contains $96\times 100$ examples of which $70\%$/$15\%$/ $15\%$ lead to the training/validation/test datasets. After training the NN (details are available in Methods and the Supplementary Information) we can estimate the Rabi frequency $\Omega_{\rm{tg}}$ of experimentally collected responses by inputting them into the NN.

\begin{table}[t]
\caption{Outputs $y_1$ obtained from the NN in Scenario i }
	\centering
	\begin{tabular}{ccc}
		\hline
		\hline
		\makecell[c]{$a_1$($\times 2 \pi$ kHz)} &\makecell[c]{$y_1$  ($\times 2 \pi$ kHz)  \\with $N_m=100$} &  \makecell[c]{$y_1$  ($\times 2 \pi$ kHz) \\  with $N_m=30$ }\\ 
		\hline
		1.1487 & 1.1827   & 1.1731   \\
		\hline
		1.7229 & 1.7473 & 1.8060 \\
		\hline
		2.2566 & 2.3109 & 2.3207 \\
		\hline
		2.8760 & 2.8616 & 2.8527 \\
		\hline
		3.4429 & 3.4961 & 3.4947 \\
		\hline
		4.0098 & 4.0391 & 4.0502 \\
		\hline
		4.5778 & 4.6283 & 4.6386 \\
		\hline
		5.1834 & 5.1856 & 5.2208 \\
	        \hline
		5.7140 & 5.7448 & 5.7297 \\
		\hline
		6.2797 & 6.1482 & 6.2134 \\
		\hline
		6.8397 & 6.8358 & 6.6896 \\
		\hline
		7.3927 & 7.3086 & 7.3471 \\
		\hline
		7.9319 & 8.0864 & 8.1129 \\
		\hline
		8.4527 & 8.3775 & 8.3870 \\
		\hline
		8.9493 & 8.8414 & 8.7825 \\
		\hline
		\hline				
	\end{tabular}\\%
\justifying\vspace{2mm}
\noindent{The input data of the NNs are the experimental responses corresponding to the targets $a_1 = \Omega_{\rm{tg}}$. These comprise the  average values collected for $N_p=151$ time instants in the interval $t\in [0, t_f]$ with a finite number of shots $N_m = 100$ and $N_m = 30$.\label{Comparison-finite-shots}}
\end{table}

In particular, we harvest sensor responses for $N=15$ values of $\Omega_{\rm{tg}}$ (while we select other RF parameters as $\phi_{\rm{tg}}=0$ and $\omega_{\rm{tg}} =(2\uppi)\times 10.56$ MHz) which do not belong to the training/validation/test datasets, and for a number of shots $N_m=100$, and $N_m=30$.  
In Table \ref{Comparison-finite-shots} we list the estimations $y_1$ from the NN for each target $a_1\equiv\Omega_{\rm{tg}}$. Each output $y_1$ is obtained after feeding the NN with the experimentally acquired response consisting of  average values measured during $t \in [0, t_f]$ with $N_p=151$. We show the results for a number of shots $N_m=100$ and $N_m=30$. 
The regressions of the NN outputs with respect to $\Omega_{\rm{tg}}$ are shown in Fig.~\ref{average-times} (a, b). 

\begin{figure*}[t]
\begin{center}
\scalebox{0.27}[0.27]{\includegraphics{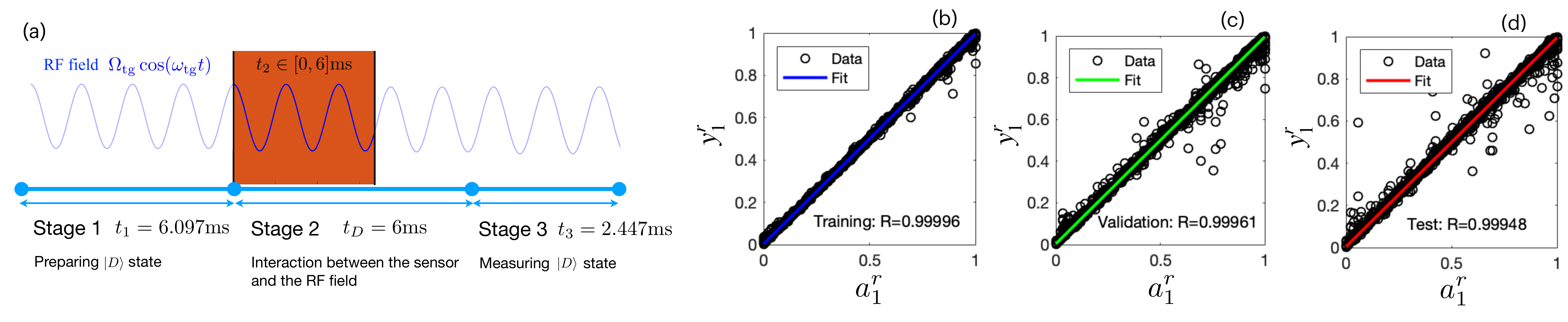}}
\caption{\label{regression-singleshot} 
Estimation results from Scenario ii, continuous data acquisition. (a) Schematic configuration of the continuous data acquisition scheme. Each binary value is obtained after completing the preparation, interaction, and measurement stages. (b-d) Regression of the rescaled outputs $y_1^r$  from the NN with respect to the targets $a_1^r = \Omega_{\rm{tg}}^r$. The regression lines $y_1^r = \alpha a_1^r + \beta$ comprise (b) $\alpha =0.9946$, $\beta = 0.0032$,  (c) $\alpha = 0.9932 $, $\beta = 0.0040$, (d) $\alpha = 0.9925 $, $\beta = 0.0041$, while the correlation coefficients $\rm{R}$ between the outputs and the targets are all larger than $0.999$. All inputs $\textbf{X}^r$, outputs $\textbf{Y}^r = \{y_1^r\}$, and  targets $\textbf{A}^r = \{a^r_1\}$ are rescaled into the range $[0,1]$.}
\end{center}
\end{figure*}

The average accuracy in the estimation of the different $\Omega_{\rm{tg}}$ is defined as $\bar{F} = \frac{1}{N} \sum_{j=1}^N  F_j $, with $F_j =1- |y_1^j - a_1^j| / a_1^j$, and $a_1 = \Omega_{\rm{tg}}$. With the values in Table~\ref{Comparison-finite-shots} we find the results $\bar{F}=98.76\%$ for $N_m=100$ with a standard deviation ($\rm{SD}$) of the $F_j$ set ${\rm SD}= 0.7762\%$. In the case of $N_m=30$, we find $\bar{F}=98.31\%$ with ${\rm SD} = 1.1483\%$. We remark that these highly accurate estimations were obtained with examples that comprise values of $\Omega_{\rm{tg}}$ leading to sensor responses that depart from the harmonic case. In particular, in Fig.~\ref{average-times} (c,d,e) we show the cases  for $\Omega_{\rm{tg}} = (2\uppi)\times 1.1487$ kHz (c) and $\Omega_{\rm{tg}} = (2\uppi)\times 8.9493$ kHz (d,e), leading to harmonic and non-harmonic responses respectively  (both cases comprise $N_m=30$). On the one hand, in  Fig.~\ref{average-times} (c) we show that the sensor response (diamonds) follows  the harmonic function $P_D(t) = \cos^2(\uppi t/t_{\rm{R}})$ (blue-solid line). On the other hand,  in Fig.~\ref{average-times} (d) one can observe that  the response (diamonds)  deviates from the harmonic case, while Fig.~\ref{average-times} (e) shows that the same response fits to a non-harmonic function (red-solid curve) obtained via numerical simulations of the Hamiltonian (see Supplementary Note 1). By introducing these two responses into our NN, we get the outputs $y_1 =(2\uppi) \times1.1731$ kHz and $y_1=(2\uppi) \times8.7825$ kHz which result in large accuracies $F = |y_1 - a_1| / a_1 = 97.84\%, 98.14\%$, respectively. In the Supplementary Information, we show further examples including both Rabi frequency $\Omega_{\rm{tg}}$ and potential detunings $\xi$ between the target frequency and the sensor hyperfine transition.

\subsection*{Scenario ii: Continuous data acquisition} 
Now we consider a scenario that comprises single-shot measurements on the ion (i.e. $N_m =1$). This scheme is relevant in situations where no reinitialization of the RF field is possible. We demonstrate that estimation of RF parameters is still feasible in these conditions. 

We get one binary value from the atomic sensor (0 or 1) after completing the three experimental stages illustrated in Fig.~\ref{regression-singleshot} (a). More specifically, at stage $1$ we cool the ion and prepare the $|D\rangle$ state in a time $t_1 = 6.097$ ms. At stage $2$ the  ion is allowed to interact with the target field for a time $t_2\in[0, 6]$ ms. At stage $3$ we readout the ion  in a time $t_3= 2.447$ ms. We repeat these three stages $251$ times (leading to a binary string comprising 251 numbers) where $t_2$ varies in the interval $[0, 6]$ ms with a step $0.024$ ms. Note that during the three stages the RF source is always on, thus to avoid potential damage on initialization and readout we switch the RF signal to a dummy load at stages 1 and 3.

Via numerical simulations we train a NN in accordance to the scheme in Fig.~\ref{regression-singleshot} (a). In particular, we use $96$ values for $\Omega_{\rm{tg}}$ in the range $(2\uppi) \times[0.5, 10]$ kHz  (with a step of $(2\uppi) \times0.1$ kHz) and repeat the data acquisition process $1800$ times for each $\Omega_{\rm{tg}}$. Thus we generate $96\times 1800 =172800$ examples of which $70\%$/$15\%$/$15\%$ are used to build the training/validation/test datasets. After training the NN, we find the regression accuracy of the training/validation/test datasets shown in Fig.~\ref{regression-singleshot} (b-d). Note that data in Fig.~\ref{regression-singleshot} (b-d) is rescaled into $[0, 1]$ as this is a standardized procedure in NNs.

Now, we experimentally obtain the sensor responses for 8 randomly chosen values of $\Omega_{\rm{tg}}$ in the range  $(2\uppi) \times[0.5, 10]$ kHz which do not belong to the training/validation/test datasets (in addition, we select  other RF parameters as $\phi_{\rm{tg}}=0$ and $\omega_{\rm{tg}} =(2\uppi)\times 10.03$ MHz). We remark that the obtained responses range from the harmonic shape to those deviating from it. After 251 measurements for each $\Omega_{\rm{tg}}$, we get binary strings including 251 numbers (0 or 1) where each number is obtained according to the scheme in Fig.~\ref{regression-singleshot} (a). When inputting  each string into the trained NN, we get one output $y_1$ from the NN. In order to study the stability of the NN prediction, we have repeated 20 times the data acquisition for each $\Omega_{\rm{tg}}$. In Table.~\ref{Comparison-singleshot} we show the average value $\bar{y}_1$ of the results from the NN and the standard deviation (\rm{SD}) based on $20$ experimentally obtained strings. 

\begin{table}[h]
\caption{\label{Comparison-singleshot}Estimation results from the NN in Scenario ii}
	\centering
	\begin{tabular}{ccc}
		\hline
		\hline
		\makecell[c]{Targets\\$a_1$($\times 2 \pi$ kHz)} &\makecell[c]{ Average values\\$\bar{y}_1$  ($\times 2 \pi$ kHz) } &  \makecell[c]{ Standard deviation \\SD  ($\times 2 \pi$ kHz)} \\
		\hline
		0.7542  & 0.8417 & 0.0690  \\
		\hline
		1.1840&  1.2759 & 0.0833   \\
		\hline
		1.4044  &1.4384 & 0.0222  \\		
		\hline
		1.6206 & 1.6542   & 0.0660   \\
		\hline
		2.1572  & 2.1720 & 0.0543  \\		
		\hline
		4.2265 & 4.2761 & 0.0594  \\		
		\hline
		6.3960 & 6.2988 & 0.0776 \\	
		\hline
		8.3689 & 8.2255 & 0.1531 \\	
		\hline
		\hline				
	\end{tabular}\\%
\justifying\vspace{2mm}
\noindent{\label{Comparison-singleshot}Average value of the estimation  $\bar{y}_1$ provided by the NN and its corresponding standard deviation based on 20 experimental acquisitions.}
\end{table}

\begin{figure}[t]
\begin{center}
\scalebox{0.2}[0.2]{\includegraphics{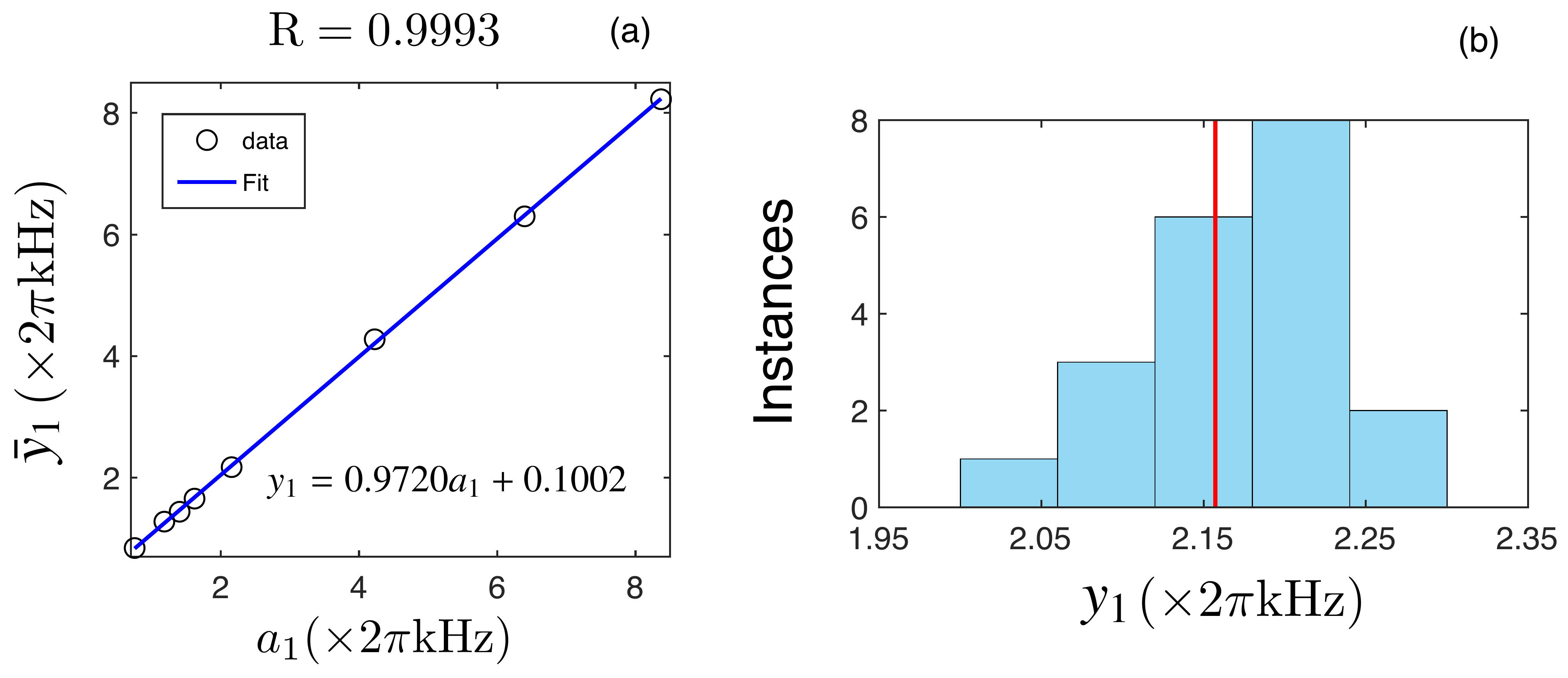}}
\caption{\label{single-shot} Acccuracy analysis for continuous data aqcuisition. 
(a) Regression of $\bar{y}_1$ obtained from the NN with respect to each target. In addition, the fit line (blue) and the correlation coefficient $\rm{R}$ are shown. 
(b) Histogram of 20 outputs $y_1$ from the NN for the case $a_1  = (2\uppi
) \times 2.1572 $kHz denoted with a vertical red line.}
\end{center}
\end{figure}

In addition, the regression of the outputs $\bar{y}_1$ with respect to the targets  is in Fig. \ref{single-shot} (a). 
Finally, in Fig. \ref{single-shot} (b) we illustrate for $a_1 = (2\uppi)\times 2.1572$ kHz the histogram of the NN outputs $y_1$ obtained after feeding the NN with $20$ strings of experimental data. The above analysis illustrates the ability of  NNs to achieve accurate estimations in scenarios involving single-shot measurements (thus, when the RF field is not controllable) leading to highly versatile quantum sensors.

\section*{Discussion}
One can resort to other estimators for predicting parameters, e.g., using Bayesian inference. Following the well-known Bayes theorem, one may compute the posterior distribution $p(\Theta|\textbf{X}) \propto p(\textbf{X}|\Theta) p(\Theta)$ where $p(\Theta)$, $p(\textbf{X}|\Theta)$ denote the prior and likelihood respectively, while $\textbf{X}$ refers to the data obtained by interrogating the quantum sensor at different time instances, and $\Theta =\{\theta_1, ..., \theta_k \}$ denotes $k$ unknown parameters which we aim to estimate by our quantum sensor. For a Bayesian estimator, an accurate microscopic model is needed in order to calculate the likelihood $p(\textbf{X}|\Theta)$. More specifically, this is 
\begin{eqnarray}
& p(\textbf{X}|\Theta)=\Pi_{i=1}^{N_p}f(X_i, N_m, \tilde{P}_i(t_i; \Theta))
\nonumber
\\
\textrm{with}   & f(x, n, p) = \frac{n!}{x! (n-x)! p^x (1-p)^{(n-x)}},
\end{eqnarray}
where the function $f(x, n, p)$ refers to the probability of observing $x$ success outcomes through $n$ trials from the Binomial distribution with success probability $p$. $\tilde{P}_i(t_i; \Theta)$ is the  survival probability $P_D$ computed using the total Hamiltonian $H$ (see Supplementary Note 1: Eq. (3)) at time $t_i$ whose obtention requires to load a string of values for the $\Theta$ parameters in the microscopic model and then compute its dynamical evolution. Note this is a procedure that has to be repeated for each value of the  $\Theta$ parameters.
Finally, via the marginal distribution $p(\theta_j | \textbf{X}) = \int \Pi_{i\neq j} d\theta_i p(\Theta | \textbf{X})$, one could derive the average value and standard deviation as
\begin{eqnarray}
& \theta_j^{\textrm{est}} = \int d\theta_j \theta_j p (\theta_j | \textbf{X}),
\nonumber
\\
&(\delta\theta_j^{\textrm{est}})^2 = \int d\theta_j (\theta_j - \theta_j^{\textrm{est}})^2 p (\theta_j | \textbf{X}).
\end{eqnarray}

In contrast, less prior knowledge of the microscopic model is needed when using NNs. This owes to NNs learn the input-output relation from the training/validation/test datasets that can be obtained from 
numerical simulations, or directly from experiments. The latter is especially useful when numerical simulation of the system dynamics becomes challenging (for instance, in a sensor that consists on several entangled ions).

In this manner, our work demonstrates the good performance in parameter estimation that results from an appropriate hybridization of machine learning tools with quantum sensing techniques. This is a strategy that can be easily extended to other quantum platforms such as, e.g., nitrogen vacancy (NV) centers in diamond, to decipher complex NV responses emerging from dense nuclear samples comprising nuclear spins which are strongly coupled to the sensor and/or among them.

\section*{Methods}
\subsection*{Neural-Network-based magnetometer}
A NN enables to find the relation between $N_p$ measured data $\textbf{X} = \{x_1, x_2, ..., x_{N_p}\}$, and $n$ output data $\textbf{Y} = \{y_1, y_2, ...y_n\}$ that  approach the targets $\textbf{A}= \{a_1, a_2, ...,a_n\}$. During the training stage of the NN, the following cost function
\begin{equation}
C = \frac{1}{nN} \sum_{j=1}^N \sum_{i=1}^n (y_i^j -a_i^j)^2
\end{equation}
is minimized for a training set that comprises $N$ examples. This is done by using gradient descent methods such that the NN parameters (i.e., weights and biases) are adjusted to satisfy $F(\textbf{X}) = \textbf{Y} \approx \textbf{A}$. 

In our case, we deal with an $^{171}$Yb$^{+}$-magnetometer where we aim to estimate RF parameters from experimentally collected responses by inputting them into the NN.
The input data contains, in  Scenario i, average values obtained from a reduced number of measurements and, in Scenario ii, a sequence containing binary values continuously acquired from single-shot measurements.
For the first case, the input data string $\textbf{X}=\{P_1, P_2, ..., P_{N_p}\}$ consist on the  average values $P_i$ ($i\in[1, N_p]$) collected in a time interval $[0, t_f]$ for a specific set of targets $\textbf{A}$. The simulated average value is $P_i  = \sum_{n=1}^{N_m} z_n^i / N_m$ for a number of shots $N_m$, where the binary outcome is drawn from a Bernoulli distribution $z_n^i \sim B(1, P_D(t_i)) \in\{0, 1\}$. In the second scenario that comprises continuous data acquisition, the input data string $\textbf{X}$ is made of binary numbers, $0$ and $1$, which are obtained according to the scheme in Fig. \ref{regression-singleshot} (a). Repeating this procedure $N$ times, we achieve the whole dataset that comprises $N$ examples.
In both cases, the examples with the data strings \textbf{X}, $\textbf{Y}$, and $\textbf{A}$ are computed by selecting a number of values of the targets in, and beyond, the regime leading to harmonic sensor responses. Among all the examples of the total datasets, $70\%/15\%/15\%$ lead to the training/validation/test sets. A number of repetitions for each data acquisition is repeated such that the NN learns the statistical fluctuations resulted from a reduced number of measurements,

\subsection*{Experimental timing sequence}
\begin{figure}[t]
\begin{center}
\scalebox{1}[1]{\includegraphics{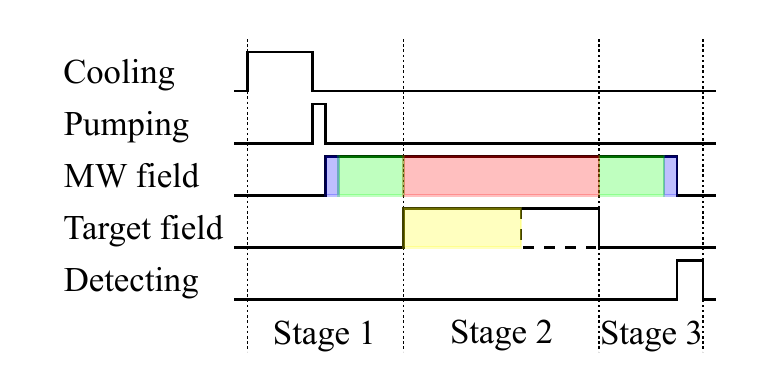}}
\caption{\label{sequence_diagram}The sequence diagram of the experiment. The whole cycle could be divided into three stages. Stage 1 consists of cooling, initialization and preparation of the state $|1\rangle$ (the first blue zone) and the state $|D\rangle$ (the first green zone). At Stage 2 (the pink zone of legth $t_D$) we allow the ion interact with the target field at the interval $[0, t_2]$ (the yellow zone), where $t_2 \leq t_D$. In stage 3, we transfer the state $|D\rangle$ back to $|0\rangle$ and detect the probability of $|0\rangle$.}
\end{center} 
\end{figure}
As shown in Fig. \ref{sequence_diagram}, in each cycle (each period to obtain one value $P_i$ ($i \in [1, N_P]$)  of a response / input data string), the ion is cooled down approximately in the Doppler limit by a red-detuning 369.5 nm laser starting from stage 1. After that, the state of the ion is initialized to $|0\rangle$ by an optical pumping process. After a 0.5 $\rm{\mu s}$ 'MW trigger' signal, a microwave $\uppi$-pulse resonant with  $|0 \rangle \leftrightarrow |1\rangle$ transfers the state from  $|0 \rangle$ to $|1\rangle$. Subsequently, the STIRAP pulses drive the system from the state  $|1\rangle$ to the dark state $|D\rangle$. At stage 2, the amplitudes of MW fields are held at a constant $\Omega$ within the fixed time interval $[0, t_D]$. Simultaneously, the target RF field is applied for the time interval $[0, t_2]$, where $t_2 \leq t_D$. For the scheme of a finite number of measurement (Scenario i), after the time instant $t_2$ the RF field is removed by turning off the AWG2 output. The RF target field is restarted at the next cycle such that the repetition of measurement can be done. For the scheme of single-shot measurement, Scenario ii, the RF signal is switched to a dummy load after the time instant $t_2$ while the origin RF source is always on. At stage 3, the STIRAP pulses transfer the rest of population of $|D\rangle$ back to $|1\rangle$.
Another  $\uppi$-pulse transfers the population of state  $|1\rangle$ to  $|0\rangle$. Thus, we can get the response of $|D\rangle$ by measuring the probability of $|0\rangle$, and a state dependent fluorescence detection could be used to determine it. 
Before and after the cycle, there are extra 'idle' steps (1 $\rm{\mu s}$) needed to end this cycle and start the next cycle.
In our experiments, the timings are all controlled by a TTL pulse generator (Spincore PB24-100-4k-PCIe) see TABLE~\ref{Sequence} for the specific values of the time intervals invested on each process.

\begin{table}[h]
	\caption{\label{Sequence} Experimental timings for two scenarios ('Idle' and 'MW trigger' are not shown)}
	\centering
	\begin{tabular}{lcc}
		\hline
		\hline
		\makecell[c]{} &\makecell[c]{Scenario i} &  \makecell[c]{Scenario ii} \\
		\hline
		Cooling & 2 ms & 4 ms \\
		\hline
		Pumping & 50 $\rm{\mu s}$ & 50 $\rm{\mu s}$ \\
		\hline
		$\pi$ - pulse & $\sim$45.45 $\rm{\mu s}$ & $\sim$45.45 $\rm{\mu s}$ \\
		\hline
		$|1\rangle \rightarrow |D\rangle$ & 2 ms & 2 ms \\
		\hline
		Sensing window & 2.828 ms & 6 ms \\
		\hline 
		$|D\rangle \rightarrow |1\rangle $ & 2 ms & 2 ms \\
		\hline
		$\pi$ - pulse & $\sim$45.45 $\rm{\mu s}$ & $\sim$45.45 $\rm{\mu s}$ \\		
		\hline
		Detecting & 400 $\rm{\mu s}$ & 400 $\rm{\mu s}$ \\
		\hline
		\hline				
	\end{tabular}
\end{table}	


\section*{Data availability}
\noindent The authors declare that the data supporting the findings of this study are available within the article and its Supplementary Information Files. Extra data are available from the corresponding author upon request.

\section*{acknowledgments}
\noindent  This work was supported by the National Key Research and Development Program of China (No. 2017YFA0304100), National Natural Science Foundation of China (Nos. 11774335, 11734015), the Key Research Program of Frontier Sciences, CAS (No. QYZDY-SSWSLH003), Innovation Program for Quantum Science and Technology (Nos. 2021ZD0301604, 2021ZD0301200). Y.C. acknowledges to the support of Students' Innovation and Entrepreneurship Foundation of USTC. This work was partially carried out at the USTC Center for Micro and Nanoscale Research and Fabrication.
Y.B. acknowledges to the EU FET Open Grant Quromorphic (828826), the QUANTEK project (ELKARTEK program from the Basque Government, expedient no. KK-2021/00070), the project ``BRTA QUANTUM: Hacia una especialización armonizada en tecnologías cuánticas en BRTA'' (expedient no. KK-2022/00041).
J.~C. acknowledges the Ram\'{o}n y Cajal   (RYC2018-025197-I) research fellowship, the financial support from Spanish Government via EUR2020-112117 and Nanoscale NMR and complex systems (PID2021-126694NB-C21) projects, the EU FET Open Grant Quromorphic (828826),  the ELKARTEK project Dispositivos en Tecnolog\'i{a}s Cu\'{a}nticas (KK-2022/00062), and the Basque Government grant IT1470-22.

\section*{Competing interests}
\noindent The authors declare no competing interests.

\section*{Author contributions}
\noindent Y.C and Y. B. are the co-first authors. Y.C., R.H, Y.-F. H. performed the experimental measurements. Y.B. and J.C. developed the theoretical model and analyzed the data. All authors drafted the work, wrote the manuscript and approved the completed version.


\pagebreak
\widetext
\begin{center}
\textbf{ \large Supplemental Information for \\ A Neural Network Assisted  $^{171}$Yb$^{+}$ Quantum Magnetometer}
\end{center}

\setcounter{equation}{0} \setcounter{figure}{0} \setcounter{table}{0}
\makeatletter 
 \global\long\def\thefigure{S\arabic{figure}}
 \global\long\def\bibnumfmt#1{[S#1]}
 \global\long\def\citenumfont#1{S#1}
\renewcommand{\thesection}{\Alph{section}}
\numberwithin{equation}{section}

\section{ Supplementary Note 1: Hamiltonian}\label{sec:H}
For an  $^{171}$Yb$^{+}$ ion in an external static magnetic field  $B_z$, external drivings $ B_x^j \cos{(\omega_j t +\phi_j)}$ can be applied, leading to the Hamiltonian 
\begin{equation}
H = A \ {\bf J} \cdot {\bf I}  + \gamma_e B_z J_z - \gamma_n B_z I_z + \sum_j(\gamma_e B_x^j  J_x - \gamma_n B_x^j I_x) \cos{(\omega_j t + \phi_j)},
\end{equation}
where $A \approx (2\uppi) \times 12.643$ GHz, the electronic/nuclear gyromagnetic ratios are $\gamma_e = (2\uppi)\times 2.8024$ MHz/G and  $\gamma_n\equiv \gamma_{^{171}\rm Yb^{+}} = (2\uppi)\times 4.7248$ kHz/G.  ${\bf J}$ and ${\bf I}$ are electron and nuclear spin-1/2 operators that can be written in a basis $\{ |1 1\rangle, |1 0\rangle, |0 1\rangle, |0 0\rangle\}$ such that $J_z |0 \ m\rangle  =-\frac{1}{2} |0 \ m\rangle$, $J_z  |1 \ m\rangle =\frac{1}{2} |0 \ m\rangle$ and $I_z |m \ 0\rangle =-\frac{1}{2}|m \ 0\rangle$, $I_z |m \ 1\rangle =\frac{1}{2}|m \ 1\rangle$ for $m=0,1$. 
In the basis  $\left\{ |1 \rangle, |\acute 0\rangle, |-1\rangle, | 0\rangle\right\}$ obtained after diagonalizing $A \ {\bf J} \cdot {\bf I}  + \gamma_e B_z J_z - \gamma_n B_z I_z$ we have
\begin{eqnarray}
\label{H-1}
H &=& \omega_1 |1\rangle\langle 1| + \omega_{\acute 0} |\acute 0\rangle\langle \acute 0| + \omega_{-1} |-1\rangle\langle -1| + \omega_{0} | 0\rangle\langle 0| 
+\sum_j \Omega_j \bigg[ |1\rangle\langle \acute{0}| +  |1\rangle\langle 0| +  |\acute0\rangle\langle -1| +  |0\rangle\langle -1|  + {\rm H.c.} \bigg]  \cos{(\omega_j t )} 
\end{eqnarray}
with $ \omega_{1} = \frac{A}{4} + (\gamma_e-\gamma_n) \frac{B_z}{2}$, $\omega_{-1} = \frac{A}{4} - (\gamma_e-\gamma_n) \frac{B_z}{2}$, $ \omega_{\acute{0}} \approx \frac{A}{4}  + \frac{(\gamma_e+\gamma_n)^2}{4 A} B_z^2$, and $ \omega_{0} \approx -\frac{3A}{4}  - \frac{(\gamma_e+\gamma_n)^2}{4 A} B_z^2$. 

Our scheme includes two MW driving fields ($j = 1, 2$) resonant with the $|0\rangle \leftrightarrow |1\rangle$ and $|0\rangle \leftrightarrow |-1\rangle$ hyperfine transitions with phases $\phi_1=\phi_2= 0$ and Rabi frequencies $\Omega_1 = \Omega_2$ (where $\Omega =\Omega_{1,2} = B_x^j \gamma_e / 2\sqrt{2} =  (2\uppi)\times5.5$kHz) that cancel magnetic field fluctuations. 
\begin{figure}[b]
\begin{center}
\hspace{-0. cm}\scalebox{0.25}[0.25]{\includegraphics{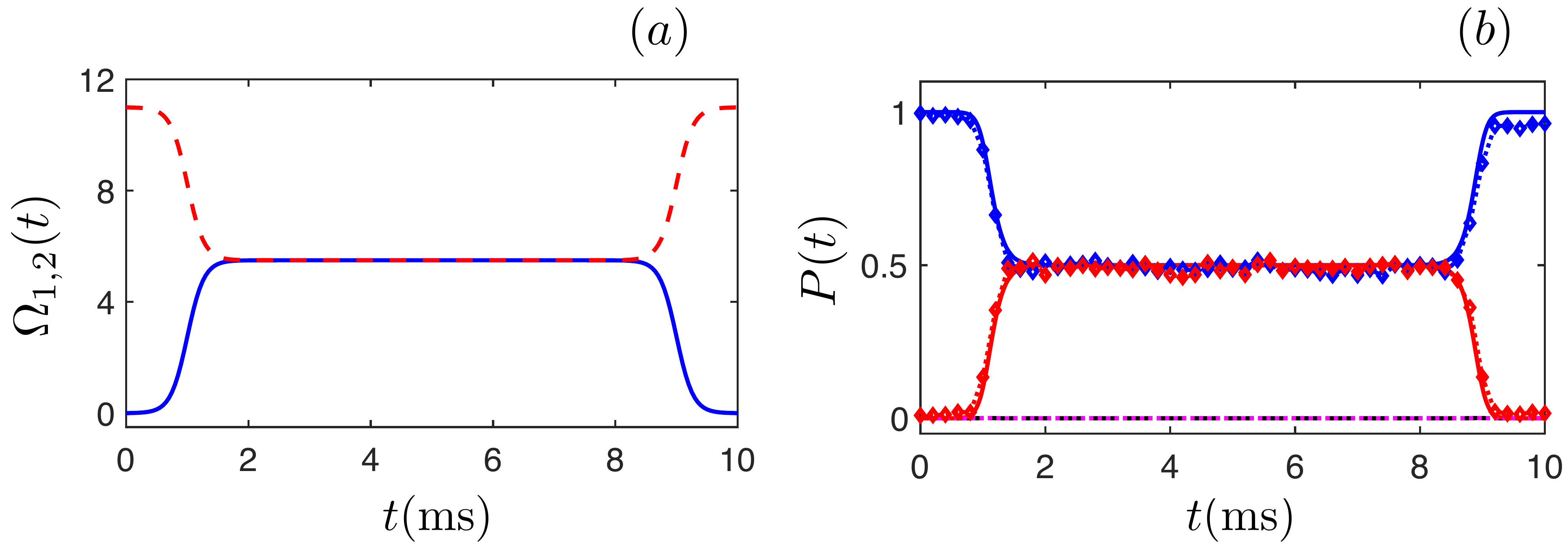}}
\caption{\label{pulses}\textbf{MW drivings and population transfer.} (a) To generate the dark state $|D\rangle$, two MW drivings $\Omega_1 (t)$ (solid blue) and $\Omega_2 (t)$ (dashed red) are applied on the $|0 \rangle \leftrightarrow |1\rangle$ and $|0\rangle \leftrightarrow |-1\rangle$ transitions. During the time interval that comprises $2 - 8$ ms we have $\Omega = \Omega_1 = \Omega_2 = 2\uppi \times 5.5$kHz. (b) Numerical simulation of population transfer of the states $|1\rangle$ (solid blue), $|-1\rangle$ (solid red) and $|\acute{0}\rangle$ (dot-dashed black), $|0\rangle$ (dotted purple), where the dark state is maintained in the interval $[2, 8]$ ms. For comparison, experimentally obtained population of the states $|1\rangle$ (blue diamond) and $|-1\rangle$ (red diamonds) are shown.
}
\end{center}
\end{figure}
When an RF target field near to the $|\acute{0}\rangle \leftrightarrow |1\rangle$ transition is included, we get the the following Hamiltonian  in the dressed state basis $|u\rangle, |d\rangle, |D\rangle, |0\rangle$
\begin{widetext}
\begin{eqnarray}
\nonumber
H(t) &=&  \frac{\Omega}{\sqrt{2}} (|u\rangle \langle u| - |d\rangle \langle d|)+\left[\frac{\Omega}{4} (|u\rangle \langle \acute{0}| + |d\rangle \langle \acute{0}| )  - \frac{\Omega_{\rm{tg}}}{2 \sqrt{2}} |D\rangle \langle \acute{0}|\right] e^{- i \xi t} + \rm{H.c.}
 \\
\nonumber
&& - \left[ \frac{\Omega}{2\sqrt{2}} (|u\rangle \langle u| - |d\rangle \langle d|) + \frac{\Omega}{4}(|u\rangle \langle D| +|D\rangle \langle d|) - \frac{\Omega}{4}(|D\rangle \langle u| + |d\rangle \langle D|)
\right] e^{i \gamma_e B_z t} + \rm{H.c.}
\nonumber
\\
&& + \frac{\Omega_{\rm{tg}}}{2} \left(\frac{1}{2} |u\rangle\langle \acute{0}|  + \frac{1}{2} |d\rangle \langle \acute{0}| - \frac{1}{\sqrt{2}} |D\rangle \langle \acute{0}|\right) e^{2i (\frac{\gamma_e B_z}{2}-\frac{\gamma_e^2 }{4A} B^2_z)t} e^{i \xi t} +\rm{H.c.}
\nonumber
\\
&& + \frac{\Omega_{\rm{tg}}}{2} \left(\frac{1}{2} |\acute{0}\rangle \langle u|+ \frac{1}{2} |\acute{0}\rangle \langle d| + \frac{1}{\sqrt{2}}|\acute{0}\rangle \langle D| \right) e^{i\gamma_e B_z t} e^{i \xi t} + \rm{H.c.}
\nonumber
\\
&& + \frac{\Omega_{\rm{tg}}}{2} \left(\frac{1}{2} |\acute{0}\rangle \langle u|+ \frac{1}{2} |\acute{0}\rangle \langle d| + \frac{1}{\sqrt{2}}|\acute{0}\rangle \langle D| \right) e^{i \frac{\gamma^2_e}{2A}B^2_z t} e^{-\xi t} +\rm{H.c.},
\label{H}
\end{eqnarray}
\end{widetext}
where $\xi = \omega_{\rm{tg}} - (\omega_{1}- \omega_{\acute{0}})$ is a potential detuning between the target frequency and the sensor hyperfine transition.

\section{ Supplementary Note 2: Pulse design to generate the state $|D\rangle$. }\label{sec:pulses}
Stimulated Raman adiabatic passage (STIRAP) pulses applied on the $|0\rangle \leftrightarrow |1\rangle$ and $|0\rangle \leftrightarrow |-1\rangle$ transitions are used to create and detect the dark state $|D\rangle = (|-1\rangle -|1\rangle) / \sqrt{2}$. In particular, these are designed in the following shape 
\begin{eqnarray}
\Omega_1(t) &=& \frac{A}{2} \left[\tanh \left(\frac{t-b_1}{c}\right)+1\right] + \frac{A}{2} \left[\tanh \left(\frac{t-b_2}{c}\right)+1\right],
\nonumber
\\
\nonumber
\Omega_2(t) &=& -\frac{A}{2} \left[\tanh \left(\frac{t-b_1}{c}\right)+1\right] 
                           \\&& - \frac{A}{2} \left[\tanh \left(\frac{t-b_2}{c}\right)+1\right] + 2A
\end{eqnarray}
We may slightly tune the parameters $b_1$, $b_2$, $c$ to adjust the time duration of keeping $\Omega_1=\Omega_2$. For instance, for a process lasting $t_f = 10$ms, with $A =5.5\times 2\uppi$ kHz, $b_1=1$ms, $b_2 = 9$ms, and $c=3\uppi/A$. it takes about $2$ ms to generate the state $|D\rangle$. This is maintained from $t = 2$ ms  to $t= 8$ ms with $\Omega = \Omega_1 = \Omega_2 = 2\uppi \times 5.5$kHz. In Supplementary Figure 1 (a) we show the controls that adiabatically evolve the initial state $|\Psi(0)\rangle = |1\rangle$. The state population evolution of theoretical and experimental data is shown in Supplementary Figure 1  (b).

\section{ Supplementary Note 3: Training of the neural network and estimation results for $\Omega_{\rm{tg}}$ from a finite number of measurement}\label{sec:average}
The data generated from numerical simulations is used to establish the NN. During the training stage, all the inputs $\textbf{X}$ and the targets $\textbf{A}$ are rescaled into the range $[0,1]$ before being used in the NN (note the notation $\textbf{A}^r$ and $\textbf{X}^r$). Correspondingly, From the outputs $\textbf{Y}^r$ (also in the range $[0, 1]$) obtained from the NN, we can get the results of the estimation $\textbf{Y}$ with the real units. The training result for estimating $\Omega_{\rm{tg}}$ via  standard gradient descent are shown in Supplementary Figure 2.
The fit line of the outputs of the NN of the whole (including training/validation/test) dataset with respect to the targets is $y^r_1 = \alpha a_1^r + \beta$, where $1-\alpha \approx 10^{-5}$ and $\beta \approx 7.7\cdot 10^{-6}$, as well as the correlation coefficient $\rm{R} = 0.99999$, prove the regression with high fidelity. 

\begin{figure}[t]
\begin{center}
\scalebox{0.25}[0.25]{\includegraphics{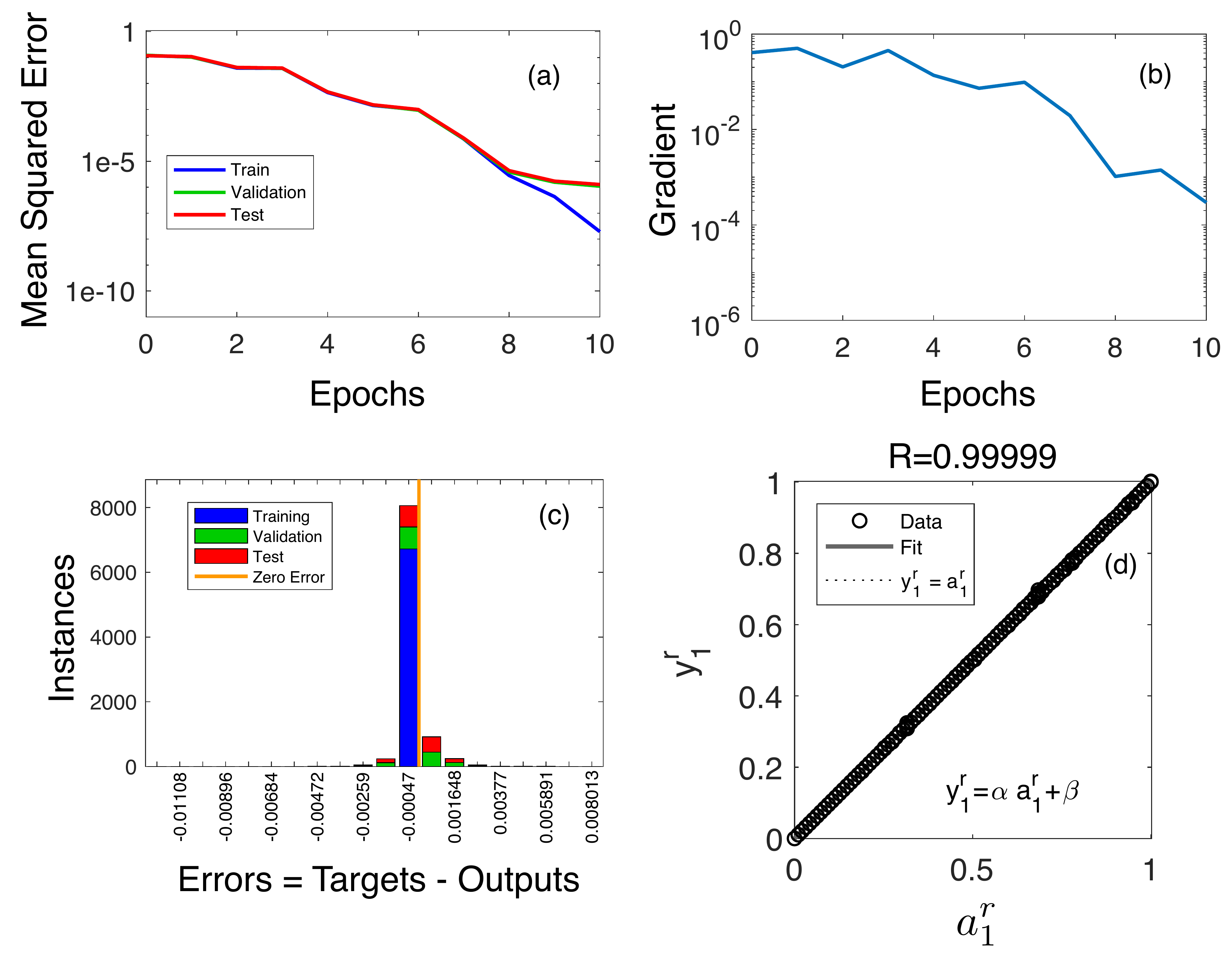}}
\caption{\label{training-average} \textbf{Training results of the NN to estimate $\Omega_{\rm{tg}}$, where the input data $\textbf{X}=\{P_1, P_2, ..., P_{N_p}\}$ is numerically derived from $96$ different values for $\Omega_{\rm{tg}} / (2\uppi) \in [0.5, 10]$ kHz with $100$ repetitions.} (a) Cost function values for training (blue) / validation (green) / test (red) datasets and (b) Gradient of the training set at each epoch. At the 10th epoch, we show the error histogram in panel (c), and (d) the comparison between the NN outputs $y_r^1$ and the targets $a_1^r$.
The fit in (d) shows the linear relation between the outputs $y_1^r$ and the target $a_1^r = \Omega_{\rm{tg}}$, i.e. $y^r_1 = \alpha a_1^r + \beta $ (solid), where $1-\alpha \approx 10^{-5}$ and $\beta \approx 7.7\cdot 10^{-6}$. The fit, the linear (ideal) case $y_1^r = a_1^r$ (dotted) almost coincide. The correlation coefficient is $\rm{R} = 0.99999$. }
\end{center}
\end{figure}

\begin{figure*}[t]
\begin{center}
\scalebox{0.4}[0.4]{\includegraphics{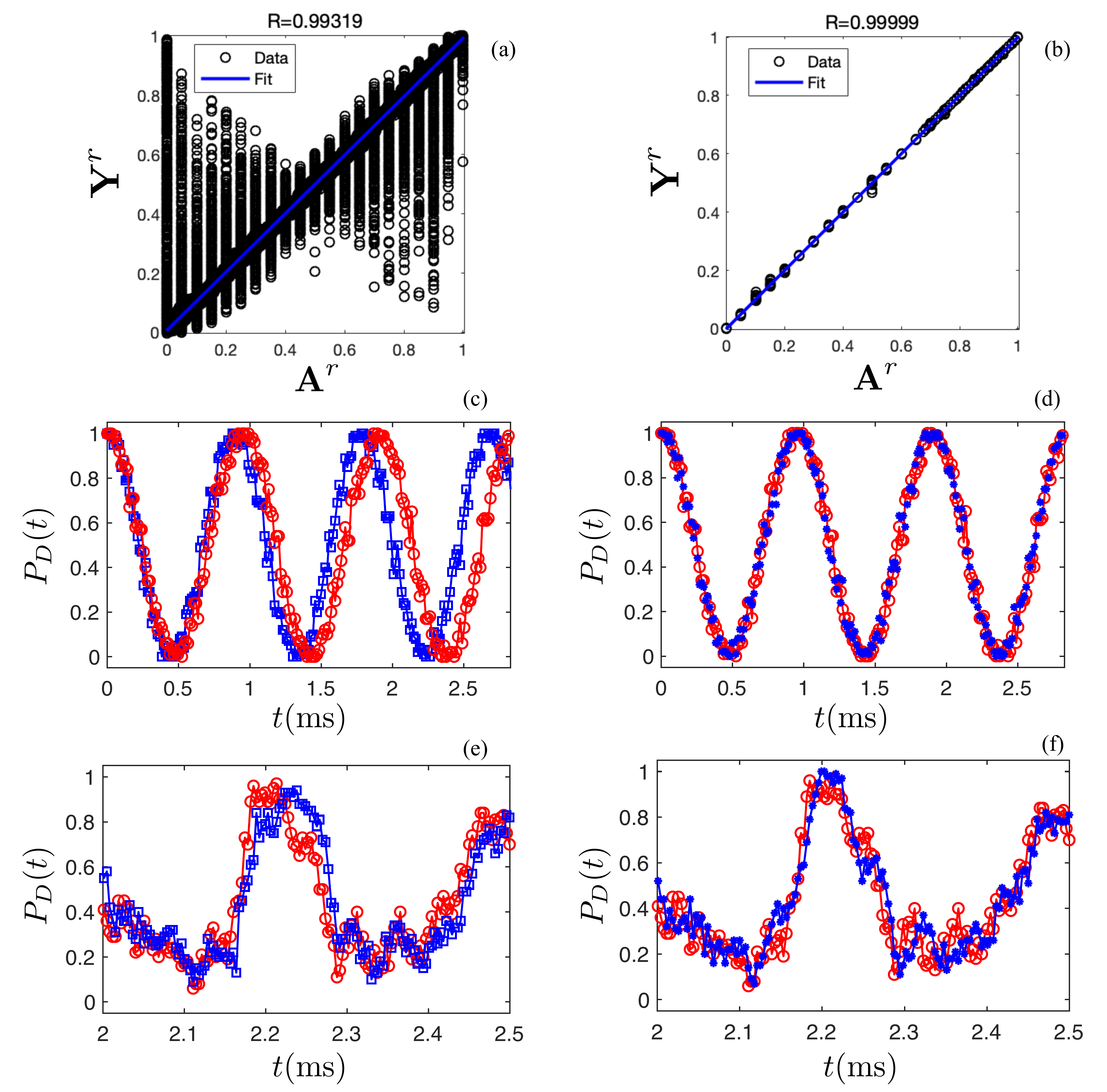}}
\caption{\label{regression-2parameters}\textbf{Estimation results for $\Omega_{\rm{tg}}$ and $\xi$ with a finite number of measurement.}
(a-b) Comparison between the rescaled outputs of the NN $\textbf{Y}=(y_1^r, y_2^r)$ and the targets $\textbf{A} = (a_1^r, a_2^r) = (\Omega_{\rm{tg}}^r, \xi^r)$, with
different range of $\Omega_{\rm{tg}}$, namely: (a) $\Omega_{\rm{tg}}^r \in [0, 1]$, (b) $[0.67, 1]$, while keeping the same detuning range $\xi^r \in [0, 1]$ (i.e. $\xi/(2\uppi) \in [-0.6, 0.6] $ kHz). All the data is renormalized to  $[0, 1]$ (note the superscript $r$).
In both plots, the fit $y_1^r = \alpha a_1^r + \beta$ and $y_2^r = \alpha a_2^r + \beta$ (solid blue)
coincide with the ideal dependences $y_1^r =a_1^r$ and $y_2^r =a_2^r$ (dashed red) for both parameters where (a) $\alpha= 0.9858$, $\beta = 0.007377$, (b) $\alpha=0.9997$, $\beta = 0.0002507$ and the correlation coefficient (a) $\rm{R} = 0.99319$, (b) $\rm{R} = 0.99999$.
(c-d) Numerical simulation of the responses $P_D(t)$ under $100$ shots. Comparisons of responses when (c) $\Omega_{\rm{tg}}/(2\uppi) =1.5$ kHz (red circle),  $\Omega_{\rm{tg}}/(2\uppi) =1.6$ kHz (blue squared) at the same $\xi = 0$, (d) $\xi =0$ (red circle),  $\xi/(2\uppi) =0.05$ kHz (blue star) at the same $\Omega_{\rm{tg}}/(2\uppi) = 1.5$ kHz, (e)  $\Omega_{\rm{tg}}/(2\uppi) =7.4$ kHz (red circle),  $\Omega_{\rm{tg}}/(2\uppi) =7.5$ kHz (blue squared) at the same $\xi = 0$, (f) $\xi =0$ (red circle),  $\xi/ (2\uppi) =0.05$ kHz (blue star) at the same $\Omega_{\rm{tg}}/(2\uppi) = 7.4$ kHz.
}
\end{center}
\end{figure*}

\section{ Supplementary Note 4: Estimation results for $\Omega_{\rm{tg}}$ and $\xi$ with a finite number of measurement}\label{sec:average-2paras}
We can also train a NN to recognize two parameters, namely potential detunings $\xi$ between the target frequency and the sensor hyperfine transition, and the Rabi frequency of the target field $\Omega_{\rm{tg}}$. In particular, we inspect two ranges for the parameters $\Omega_{\rm{tg}}$ and $\xi $: Firstly, we consider  $\Omega_{\rm{tg}} / (2\uppi) \in [0.5, 10]$ kHz and $\xi / (2\uppi) \in [-0.6, 0.6]$ kHz by extracting $96$ values of $\Omega_{\rm{tg}}$ with  a  step $0.1$ kHz and $51$ values of $\xi$ separated by $0.024$kHz. The input data $\textbf{X} = \{P_1, P_2, ..., P_{N_p}\}$ in each example consists of $N_p = 201$ values collected in the range $t \in [0, t_f]$, where $t_f = 2.828$ ms. To enable the NN learning statistical fluctuation from shot noise, $100$ repetitions are done for each numerical acquisition. 
Secondly, we keep the same range of detuning $\xi / (2\uppi) \in [-0.6, 0.6]$ kHz, and set another  range for $\Omega_{\rm{tg}} /(2\uppi) \in [6.9, 10]$ kHz. In both cases, $70\%$, $15\%$, $15\%$ of examples in each dataset mentioned above are randomly extracted to create the training/validation/test set in order to build the respective NN. The regression plots of the outputs from the above two NNs and the targets are illustrated in Supplementary Figure 3  (a-b). 

We observe that the regression relation experiences less fluctuations in the regime in which $\Omega_{\rm{tg}}$ is far away from harmonic shape. In particular, at small $\Omega_{\rm{tg}}$ shot noise hinders the identification of the sensor response, as the slight tuning in $\xi$ already mixes the responses. Meanwhile, the accuracy to estimate $\Omega_{\rm{tg}}$ is always high. This is proved in the numerical simulation in Supplementary Figure 3 (c-f): when adopting the same $\xi$, the responses obtained numerically under $100$ times of shots, can be identified in the presence of a small variation in $\Omega_{{\rm{tg}}}$ both at small values $\Omega_{\rm{tg}}$ (c) and large ones (d). Provided by the same $\Omega_{{\rm{tg}}}$ and a small variation in $\xi$, the responses show little difference at a small $\Omega_{{\rm{tg}}}$, whereas large difference at large $\Omega_{\rm{tg}}$.
Here, we choose the response with large $\Omega_{\rm{tg}}$ from experiments as test examples, when fixing $\omega_{\rm{tg}} = 10.03$ MHz and $N_m=50$. Inputing it into the NN established in the range $\Omega_{\rm{tg}} /(2\uppi) \in [6.9, 10]$ kHz and $\xi/(2\uppi) \in [-0.6, 0.6]$ kHz, we can get the results demonstrated in Supplementary Table \ref{Comparison}.

It is possible to separate the responses with different $\xi$ under shot noise by allowing for a longer evolution time and an increased number of shots. Consequently, the detection scheme via NNs can be improved. However, high accuracy comes at the cost of the increased time to generate the datasets.

\begin{table}[]
\caption{Estimation results for $\Omega_{\rm{tg}}$ and $\xi$ with a finite number of measurement.}
	\centering
	\begin{tabular}{lll}
		\hline
		\hline
		targets& $a_1/ (2\uppi)= 7.3920$ kHz  & $a_2/ (2\uppi) = -0.6$ kHz   \\
		
		outputs  & $y_1/ (2\uppi) = 7.5614$ kHz & $y_2/ (2\uppi) = -0.5865$ kHz  \\
		\hline
		targets& $a_1/ (2\uppi) = 7.3920$ kHz  & $a_2/ (2\uppi)  = 0$ kHz   \\
		
		outputs  & $y_1/ (2\uppi) =7.4520$ kHz & $y_2/ (2\uppi) = 0.019$ kHz  \\		
		\hline
		targets& $a_1/ (2\uppi) = 7.3920$ kHz  & $a_2/ (2\uppi)  = 0.6$ kHz   \\
		
		outputs  & $y_1/ (2\uppi) =7.4575$ kHz & $y_2/ (2\uppi) = 0.3969$ kHz  \\		
		\hline
		\hline		
	\end{tabular}\\%
\justifying\vspace{2mm}
\noindent{\label{Comparison}Comparison of the outputs $\textbf{Y} = \{y_1, y_2\}$ from the NN established for the range  $\Omega_{\rm{tg}} /(2\uppi) \in [6.9, 10]$ kHz and $\xi / (2\uppi) \in [-0.6, 0.6]$ kHz with the targets $\textbf{A} = \{a_1, a_2\}$, while the input data are experimentally obtained response where $a_1 = \Omega_{\rm{tg}} $ and $a_2 = \xi$. }
\end{table}


\section{ Supplementary Note 5: Precision analysis.} 
The measurement precision for a parameter $\theta$ encoded in a state $|\Psi\rangle$ is upper bounded by the quantum Fisher information (QFI) \cite{QFI} defined as 
$I_{\theta} = 4 \left[\langle \partial_\theta \Psi | \partial_\theta \Psi\rangle - |\langle \Psi | \partial_\theta \Psi\rangle|^2 \right].$
Consequently, the variance of the parameter estimator is lower bounded as  
\begin{eqnarray}\label{bound}
\Delta^2_{\theta} \geq \Delta^2 \theta(t_0)^{\rm{QFI}} = \frac{1}{N_T I_\theta(t_f)}.
\end{eqnarray}
The term at the right hand side of Eq.~(\ref{bound}) is accesible by computing the state at the final time $t_f$, where  $N_T = N_p \times N_m$ with $N_p$ the time instants at which the state is interrogated, and $N_m$ the number shots at each time instant. For instance, taking $\Omega_{\rm{tg}} = (2\uppi) \times4.2265$ kHz from the scheme of continuous data acquisition, we get the lower bound for the variance $\Delta {\Omega_{\rm{tg}}}^{\rm{QFI}} = (2\uppi)\times 0.003294$ kHz. Our NN gives $20$ results for $y_1$ based on $20$ strings of experimental responses leading to $\bar{y}_1= (2\uppi)\times 4.2761$ kHz, with  $\rm{SD}=( 2\uppi) \times0.0594$ kHz, see Table II in the main text. This proves that the estimation precision via the NN from the experimentally acquired responses is, as expected, below the limit imposed by QFI.

\end{document}